\documentclass[12pt,preprint]{article}%
\usepackage{amssymb}
\usepackage{amsmath}
\usepackage{graphics}
\usepackage{epsfig}
\usepackage{amsfonts}
\usepackage{graphicx}%
\newcommand{\eqb}{\begin{equation}}
\newcommand{\eqe}{\end{equation}}
\newcommand{\dmb}{\begin{displaymath}}
\newcommand{\dme}{\end{displaymath}}

\newcommand{\eab}{\begin{eqnarray}}
\newcommand{\eae}{\end{eqnarray}}

\newcommand{\be}{\begin{equation}}
\newcommand{\ee}{\end{equation}}

\begin{document}
\begin{titlepage}
\begin{flushright}
\end{flushright}
\vspace{0.4cm}
\begin{center}
\Large{Bursts of low-energy electron-positron pairs in TeV-range collider physics}
\vspace{1.0cm}\\
\large{Francesco Giacosa$^\dagger$ and Ralf Hofmann$^*$}
\end{center}
\vspace{1.0cm}
\begin{center}
{\em $\mbox{}^\dagger$ Institut f\"ur Theoretische Physik\\
Universit\"at Frankfurt\\
Johann Wolfgang Goethe - Universit\"at\\
Max von Laue--Str. 1\\
60438 Frankfurt, Germany}
\end{center}
\vspace{1.0cm}
\begin{center}
{\em $\mbox{}^*$ Institut f\"ur Theoretische Physik\\
Universit\"at Karlsruhe (TH)\\
Kaiserstr. 12\\
76131 Karlsruhe, Germany}
\end{center}
\vspace{1.0cm}
\begin{abstract}
In this Letter we investigate the possible emission of low-energy electron
neutrinos and electron-positron pairs of anomalously large multiplicity
in close-to-central $pp$ collisions at LHC. The scenario is
based on confining SU(2) Yang-Mills dynamics of Hagedorn temperature $\sim m_e=511\,$keV
being responsible for the emergence of the lightest lepton family and the
weak interactions of the Standard Model. Although cut off
by LHC's detectors these electrons-positron bursts would be seen
indirectly by a large defect energy and thus an anomalously strong
decrease of events with interesting high-energy secondaries for increasing $\sqrt{s}$ .
This is because the formation of superconducting (preconfining)
SU(2) hot-spots `steals' a large fraction of $\sqrt{s}$ subsequently
transferring it to a
thermal spectrum of electron neutrinos,
electrons, and positrons liberated through evaporation. We thus
propose the detection of electrons and positrons of kinetic energy
$\sim m_e$ and photons of energy $\sim 2\,m_e$.
\end{abstract}
\end{titlepage}

\noindent\emph{Introduction.} The experimental program at LHC will start soon.
In this note we wish to point out that large multiplicities in $e^{+}e^{-}$
production and a spectrum of gamma rays strongly peaked at $\sim1\,$MeV are
expected to occur based on the scenario that the emergence and the weak
interactions of electrons, positrons, and their neutrinos are due to a pure
SU(2) Yang-Mills theory of scale $\Lambda_{e}\sim m_{e}=0.511\,$%
MeV\thinspace\cite{Hofmann2005,GH2005,ZPinch,Hofmann2007,MoosmannHofmann2008}.
Being highly unconventional, this framework needs further elaboration and
tests to be viable. Nevertheless, we feel that some consequences of this
approach to electroweak physics could be of relevance in interpreting certain
experimental signatures at LHC. It is suspected that an unexpectedly low rate
of events with an energy-momentum transfer comparable to the center-of-mass
energy of colliding protons occurs.

In \cite{Hofmann2005,ZPinch,Hofmann2007} we have explained how the confining
phase of an SU(2) Yang-Mills theory of scale $\sim0.5\,$MeV provides for the
emergence of stable and instable fermionic solitons (center-vortex loops)
where the three lightest excitations are interpreted as the (Majorana)
electron-neutrino and the electron/positron, respectively. These nonlocal
excitations \cite{MoosmannHofmann2008} are liberated when cooling the
Yang-Mills system below the Hagedorn temperature $T_{{\tiny \mbox{H}}}\sim
m_{e}=0.511\,$MeV. Shortly above $T_{{\tiny \mbox{H}}}$ the theory is
preconfining \cite{Hofmann2005} with the dual massive gauge mode decoupling at
$T_{{\tiny \mbox{H}}}$. When disregarding the mixing with U(1)$_{Y}$ this
neutral vector boson is interpreted as the $Z_{0}$ resonance. Above
$T_{c}=1.14\,T_{{\tiny \mbox{H}}}$ there is a deconfining phase, and two
massive and oppositely charged vector bosons decouple at $T_{c}$. These modes
are interpreted as the $W^{\pm}$ resonances. In this scenario the existence of
the mass hierarchy $m_{Z_{0}}/m_{e}\sim m_{W^{\pm}}/m_{e}\sim m_{e}/m_{\nu
_{e}}\sim10^{5}$ is not explained by the large vacuum expectation of an
elementary and fundamentally SU(2) charged scalar field -- the Higgs field of
the Standard Model, see for instance the review of Ref. \cite{smrev} and Refs.
therein -- but is related to the large values of effective gauge couplings at
their respective phase boundaries. Notice that the dynamical (self-induced)
gauge-symmetry breaking, compare also with Refs. \cite{thooft}, is a
consequence of the nonperturbative sector of the pure SU(2) Yang-Mills theory
which, after a suitable spatial coarse-graining, manifests itself in terms of
effective and inert (non-fluctuating) scalar fields.

In this note we compute decay signatures of preconfining (superconducting)
hot-spots of Bohr-radius size as they may be generated in head-on proton
collisions at several TeV center-of-mass energy. These signatures include
temporal event shapes, evaporation times, and lepton-emission multiplicities.
In the case that the $pp$-collision products at LHC are also viewed at
energies comparable to $m_{e}$ these signatures may prove to be relevant.
Indirectly, we suspect an unexpectedly rare detection of events with large
four-momentum transfer because the above-mentioned hot-spots `steal' out of
the collision zone considerable fractions of the original center-of-mass
energy to redistribute this energy into large numbers of slow electrons and
positrons (kinetic energy $\sim m_{e}$). This is not unlike the greenhouse
effect where a high-frequency photon after traversing the glass-wall from
outside is transformed into a large number of low-frequency photons which can
not escape the house anymore. The analysis performed in the present work is
based on simple geometric considerations and on energy-conservation. We work
in natural units where the speed of light in vacuum and Boltzmann's constant
are set equal to unity.

\noindent\emph{Size, lifetime, and decay products of superconducting
hot-spots.} For $T\sim T_{{\tiny \mbox{H}}}$ the energy density $\rho
_{{\tiny \mbox{H}}}$ in the preconfining phase is given as $\rho
_{{\tiny \mbox{H}}}=4\pi\Lambda^{3}T_{{\tiny \mbox{H}}}$ where $\Lambda\sim
m_{e}$ \cite{Hofmann2005,GH2005} is the Yang-Mills scale of the SU(2) gauge
theory. Notice that $\rho_{{\tiny \mbox{H}}}$ solely is due to the thermal
ground state in that phase because the massive, dual gauge mode decouples at
$T_{{\tiny \mbox{H}}}$. To parameterize an uncertainty in the relation between
$\Lambda$ and $m_{e}$ we introduce the dimensionless parameter $y$ as
$\Lambda=y\,m_{e}$. As a consequence, $T_{{\tiny \mbox{H}}}$ can be expressed
as $T_{{\tiny \mbox{H}}}=\frac{11.24}{2\pi}y\,m_{e}$ \cite{GH2005}, and we
have $\rho_{{\tiny \mbox{H}}}=22.48\,y^{4}m_{e}^{4}$. Moreover, we denote by
$\delta$ the fraction of the total center-of-mass energy $\sqrt{s}\sim14\,$TeV
used up by the formation of a spherical SU(2)$_{e}$ hot-spot (preconfining
phase) of initial radius $R_{0}$. Assuming the hot-spot to be homogeneously
thermalized\footnote{In reality, the radius of the hot-spot will undergo
complicated oscillations since at $T_{{\tiny \mbox{H}}}$ the pressure
$p=p_{{\tiny \mbox{H}}}=-\rho_{{\tiny \mbox{H}}}$ is maximally negative, thus
shrinking the hot-spot, while the pressure $p$ becomes positive just above the
deconfining temperature $T_{c}=1.14\,T_{{\tiny \mbox{H}}},$ thus expanding the
hot-spot. This situation should be describable by ideal hydrodynamics.
However, since the spread of the associated temperature is small, we neglect
this effect in the following and assume homogeneous thermalization at
$T_{{\tiny \mbox{H}}}.$} at $T_{{\tiny \mbox{H}}},$ the initial radius $R_{0}$
calculates as
\begin{equation}
\delta\sqrt{s}=\rho_{{\tiny \mbox{H}}}\frac{4}{3}\pi R_{0}^{3}%
\ \ \ \Rightarrow\ \ \ R_{0}=\left(  \frac{3}{4\pi}\frac{\delta\sqrt{s}}%
{\rho_{{\tiny \mbox{H}}}}\right)  ^{1/3}\,.
\end{equation}
The left panel in Fig.\thinspace1 depicts the dependence of $R_{0}$ (in units
of the Bohr radius $a_{{\tiny \mbox{Bohr}}}=5.29\times10^{-11}\,$m) on
$\delta$ for $y=1/2$ and $y=1$, corresponding to $T_{{\tiny \mbox{H}}%
}=0.89\left.  m_{e}\right.  $ and $T_{{\tiny \mbox{H}}}=1.79\left.
m_{e}\right.  $, respectively. Notice that on the scale of the proton radius
the initial hot-spot radius $R_{0}$ is gigantic for $\delta\sim O(1)$.%
\begin{figure}
[ptb]
\begin{center}
\includegraphics[
height=2.1517in,
width=5.4414in
]%
{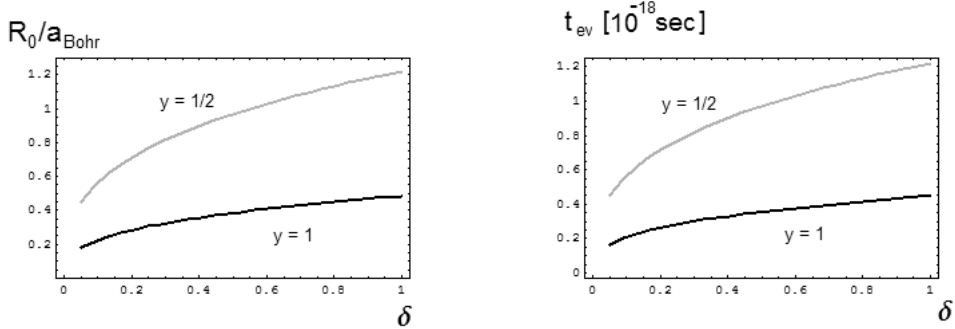}%
\caption{Left panel: Initial hot-spot radius $R_{0}$ in units of the Bohr
radius as a function of $\delta$ for $y=1/2$ (gray curve) and $y=1$ (black
curve). Right panel: Evaporation time $t_{ev}$ in units of $10^{-18}%
$\thinspace sec as a function of $\delta$ for $y=1/2$ (gray curve) and $y=1$
(black curve).}%
\label{fig1ab}%
\end{center}
\end{figure}

Once it has formed, the hot-spot starts to decay by the emission of single and
self-intersecting center vortex loops from its surface. The latter defines the
spatial location where the Hagedorn transition towards confining dynamics
takes place (traversed from inside to outside). Asymptotically far away from
the hot-spot's surface only (quasi)stable solitons, see also
\cite{MoosmannHofmann2008}, survive. These mainly are $\nu_{e}$'s and $e^{+}%
$-$e^{-}$ pairs roughly being thermalized at $T_{{\tiny \mbox{H}}}$. Close to
the surface the plasma is highly turbulent \cite{ZPinch,Hofmann2007} because
of the production and decay of unstable fermions with masses $m_{n}\sim n\cdot
m_{e}$ with $n=2,3,...$ into $\nu_{e}$ and $e^{+}$-$e^{-}$ pairs, and,
occasionally, into 1\thinspace MeV-$\gamma$'s.

For the hypothetical situation that all fermions are thermalized at a
temperature $T$ the energy density $\varepsilon_{n}$ belonging to fermions of
mass $m_{n}$ is given as
\begin{equation}
\varepsilon_{n}(T)\equiv\int_{0}^{\infty}dp\,u_{n}(p,T)\equiv\int_{0}^{\infty
}dp\,p^{2}\frac{M_{n}}{2\pi^{2}}\frac{\sqrt{p^{2}+m_{n}^{2}}}{e^{\beta
\sqrt{p^{2}+m_{n}^{2}}}+1}%
\end{equation}
where $M_{0}=2$ (Majorana neutrinos, only spin degeneracy \cite{majorananu}),
$M_{1}=2\cdot2=4$ for $e^{+}$ and $e^{-}$ (spin and charge degeneracy), and
$\beta\equiv1/T$. To evaluate the emissive power $J_{n}(T)$ (energy emitted
per unit time and unit surface) of fermions with mass $m_{n}$ we multiply the
spectral density $u_{n}(p,T)$ with $v_{n}/4$ \cite{book}. Here the velocity
$v_{n}=v_{n}(p)$ is given as $v_{n}=p/\sqrt{p^{2}+m_{n}^{2}}$. Thus, we have
\begin{equation}
J_{n}(T)=\int_{0}^{\infty}dp\,\frac{v_{n}}{4}u_{n}(p,T)=\int_{0}^{\infty
}dp\,\frac{M_{n}}{8\pi^{2}}\frac{p^{3}}{e^{\beta\sqrt{p^{2}+m_{n}^{2}}}+1}\,.
\end{equation}
The total emissive power is then given as $J(T)=\sum_{n=0}^{\infty}J_{n}(T)$.
A similar series was studied in \cite{ZPinch} demonstrating its asymptotic
nature. The latter, in turn, was shown to be responsible for the emergence of
non-equilibrium behavior (imaginary part in the pressure) by analytically
continuing the Borel resummed series back to physical `temperatures'
\cite{Hofmann2007}. Since we are only interested in what is observed far from
the emitting surface we safely may assume a thermal freeze-out at
$T_{{\tiny \mbox{H}}}$ of the (quasi)stable excitations $\nu_{e}$ and $e^{+}%
$-$e^{-}$. Thus $J(T_{{\tiny \mbox{H}}})=J_{0}(T_{{\tiny \mbox{H}}}%
)+J_{1}(T_{{\tiny \mbox{H}}})$.

Let us now turn to the time-evolution of the hot-spot radius $R=R(t)$. To do
this, recall that we assume the hot-spot to be homogeneously thermalized at
$T_{{\tiny \mbox{H}}}$. Seen from far away, the hot-spot evaporates at a
temperature $T_{{\tiny \mbox{H}}}$ into $\nu_{e}$ and $e^{+}$-$e^{-}$ where
$R(t_{0}=0)=R_{0}$ as evaluated above. By energy conservation, the
differential equation describing this evaporation reads as
\begin{equation}
\frac{dE}{dt}=\frac{d}{dt}\left(  \frac{4}{3}\pi R^{3}\rho_{{\tiny \mbox{H}}%
}\right)  =-J(T_{{\tiny \mbox{H}}})\cdot4\pi R^{2}%
\end{equation}
where $E=\frac{4}{3}\pi R^{3}\rho_{{\tiny \mbox{H}}}$ denotes the total energy
contained in the hot-spot at time $t$. The solution is easily obtained:
\begin{equation}
R(t)=R_{0}-\frac{J(T_{{\tiny \mbox{H}}})}{\rho_{{\tiny \mbox{H}}}}\,t\,.
\end{equation}
Thus, the time $t_{ev}$ needed for complete evaporation of the hot-spot reads
\begin{equation}
t_{ev}=\frac{\rho_{{\tiny \mbox{H}}}}{J(T_{{\tiny \mbox{H}}})}R_{0}\,.
\end{equation}
In the right panel of Fig.\thinspace\ 1 the quantity $t_{ev}$ is plotted as a
function of $\delta$ for $y=1/2$ and $y=1$ indicating that $t_{ev}$ is of
order $10^{-18}$ sec. Thus an initially large hot-spot (Bohr-radius size)
exhibits a decay time typically observed in electromagnetic decays of hadrons!

Let us now evaluate the number of fermions emitted. We denote by $\eta_{n}$
the number of fermions of mass $m_{n}$ emitted per unit time and surface, and
we have
\begin{equation}
\eta_{n}(T)=\int_{0}^{\infty}dp\,\frac{v_{n}}{4}\frac{u_{n}(p,T)}{\sqrt
{p^{2}+m_{n}^{2}}}=\int_{0}^{\infty}dp\,\frac{M_{n}}{8\pi^{2}\sqrt{p^{2}%
+m_{n}^{2}}}\frac{p^{3}}{e^{\beta\sqrt{p^{2}+m_{n}^{2}}}+1}\,.
\end{equation}
As before, we may restrict to the cases $n=0$ and $n=1$ after thermal
freeze-out. The total numbers $N_{\nu_{e}}$ and $N_{e^{+}-e^{-}}$ of neutrinos
and $e^{+}$, $e^{-}$ emitted are given as
\begin{equation}
N_{\nu_{e}}=\int_{0}^{t_{ev}}dt\,n_{\nu_{e}}(t)\,,\ \ \ \ \ n_{\nu_{e}%
}(t)\equiv4\pi\eta_{0}(T_{{\tiny \mbox{H}}})\,R(t)^{2}\,,
\end{equation}%
\begin{equation}
N_{e^{+}-e^{-}}=\int_{0}^{t_{ev}}dt\,n_{e^{+}-e^{-}}(t)\,,\ \ \ \ \ n_{e^{+}%
-e^{-}}(t)\equiv4\pi\eta_{1}(T_{{\tiny \mbox{H}}})\,R(t)^{2}\,.
\end{equation}
The quantities $\eta_{0}(T_{{\tiny \mbox{H}}})$ and $\eta_{1}%
(T_{{\tiny \mbox{H}}})$ are constant implying that $N_{\nu_{e}}=c\,\eta
_{0}(T_{{\tiny \mbox{H}}})$ and $N_{e^{+}-e^{-}}=c\,\eta_{1}%
(T_{{\tiny \mbox{H}}})$ with the universal constant $c$ defined as
$c\equiv4\pi\int_{0}^{t_{ev}}dt\,R(t)^{2}$. In the left panel of
Fig.\thinspace2 the numbers $N_{\nu_{e}}$ and $N_{e^{+}-e^{-}}$ are shown as
functions of $\delta$ for $y=1/2$. Notice that for $\delta=O(1)$ a very large
number of fermions is emitted and that the dependence on $\delta$ is rather
weak. Finally, in the right panel of Fig.\thinspace2 the emission rates
$n_{\nu_{e}}(t)$ and $n_{e^{+}-e^{-}}(t)$ are plotted for $y=\delta=1/2$.
Notice that the burst is particularly strong just after formation with rapidly
decreasing rates thereafter. The situation does not change qualitatively when
varying $y$ within $0.1\leq y\leq1$.%
\begin{figure}
[ptb]
\begin{center}
\includegraphics[
height=2.041in,
width=5.4423in
]%
{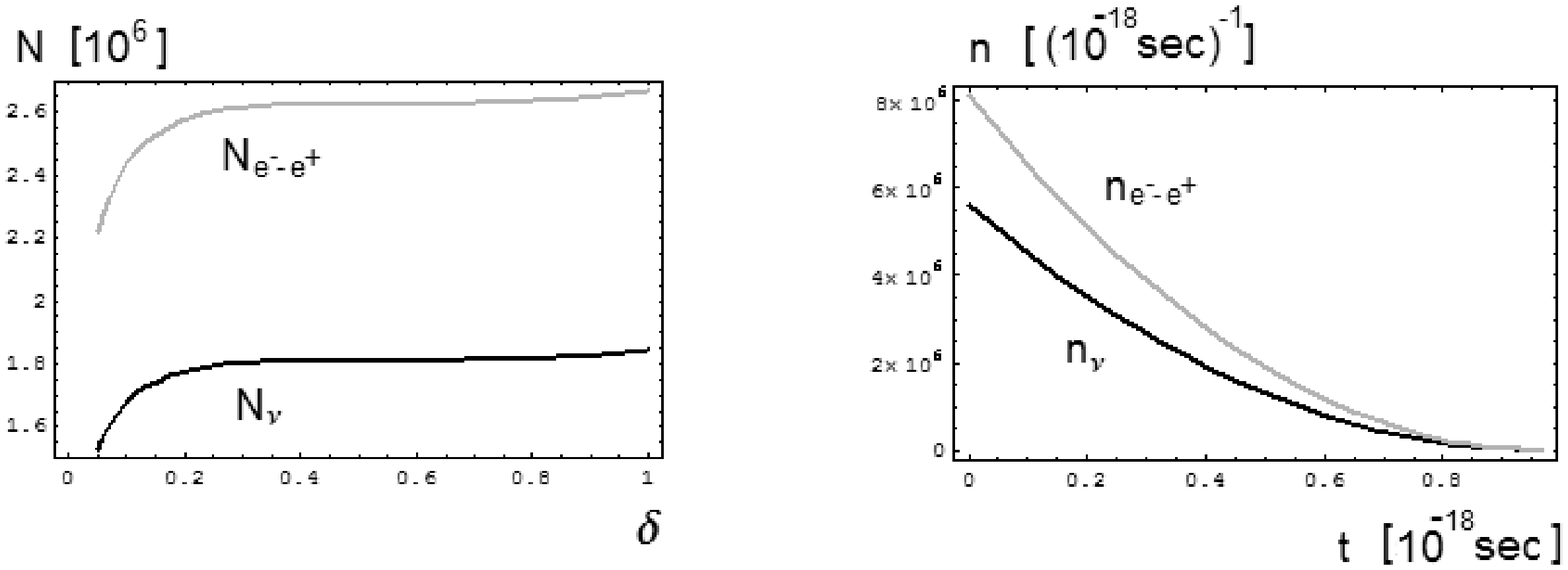}%
\caption{Left panel: number $N_{\nu_{e}}$ of neutrinos emitted (black curve)
and number $N_{e^{-}-e^{+}}$ electrons and positrons emitted (gray curve) as
functions of $\delta$ for $y=1/2$. Right panel: Emission rates for neutrinos
(black) and for electrons/positrons (grey) for $y=1/2$ and $\delta=1/2$.}%
\label{fig21b}%
\end{center}
\end{figure}

\emph{Discussion and Conclusion.} In this Letter we have calculated the
initial radius, the lifetime, the emission rates and total numbers of
neutrinos and electrons/positrons emitted by superconducting hot-spots
generated at LHC energies. These hot-spots represent the preconfining thermal
ground state in pure SU(2) Yang-Mills theory with the latter assumed to be
responsible for the emergence of the lightest lepton family and the weak
interactions \cite{Hofmann2005,GH2005}. At LHC this mechanism may serve as an
energy-shifter in the sense that the creation of preconfining hot-spots
`steals' a large fraction of the initial center-of-mass energy $\sqrt{s}$ from
the $pp$ collision zone subsequently pumping it into the creation of a large
number of thermal electrons, positrons and neutrinos at a temperature
comparable to $m_{e}=0.511\,$MeV. If this mechanism would become more
effective with increasing $\sqrt{s},$ then this would imply that the number of
interesting high-energy events quickly dies off. In addition, the
superconductivity of the hot-spot `stuff' creates magnetic fields that locally
compete with external magnetic fields and thus defocus the beam in the
collision zone. By the annihilation of electron-positron pairs we also expect
a gamma spectrum which is strongly peaked at an energy of about 1\thinspace MeV.

Notice that there are two major differences between our scenario and
black-hole evaporation \cite{page}: (i) Evaporation of black hole is induced
by the universal coupling of gravity to matter, and thus all kinds of
sufficiently stable particles are expected to appear as decay products while
we only expect a large number of neutrinos, electrons and positrons plus a
spectrum of photons peaked at $\sim$ 1 MeV. (ii) The rate of particle emission
due to our preconfining hot spots is large in an initial stage just after
formation and decreases thereafter. This is opposite to what is expected
during black-hole evaporation.

Finally, let us point out that the confining phase of an SU(2) Yang-Mills
theory of scale $\sim m_{\mu}\sim200\,m_{e}\sim100\,$MeV would provide for the
emergence of the second lepton family. The triplet $Z_{0}^{\prime}$ and
$W^{\prime}_{\pm}$ of heavy vector bosons of that theory would mediate very
weak interactions. Assuming the same decoupling values $\sim10^{5}$ of the
effective gauge couplings as experimentally inferred for the SU(2) theory
discussed above, the masses within this additional triplet would be of the
order of $10^{5}m_{\mu}\sim10$\,TeV which is just within the reach of LHC.

\end{document}